# Colossal linear magnetoelectricity in polar magnet $Fe_2Mo_3O_8$

Yuting Chang[1#], Yakui Weng[2#], Yunlong Xie[3#], Bin You[1], Junfeng Wang[1], Liang Li[1], Jun-Ming Liu[4], Shuai Dong[5*] & Chengliang Lu[1*]

[1] *School of Physics & Wuhan National High Magnetic Field Centre, Huazhong University of Science and Technology, Wuhan 430074, China*

[2] *School of Science, Nanjing University of Posts and Telecommunications, Nanjing 210023, China*

[3] *Institute for Advanced Materials, Hubei Normal University, Huangshi 435001, China*

[4] *Laboratory of Solid State Microstructures, Nanjing University, Nanjing 210093, China*

[5] *Key Laboratory of Quantum Materials and Devices of Ministry of Education, School of Physics, Southeast University, Nanjing 211189, China*

**Abstract:** Linear magnetoelectric effect is an attractive phenomenon in condensed matters and provides indispensable technological functionalities. Here a colossal linear magnetoelectric effect with diagonal component $α_{33}$ reaching up to ~480 ps/m is reported in a polar magnet $Fe_2Mo_3O_8$, and this effect can persist in a broad range of magnetic field (~20 T) and is orders of magnitude larger than reported values in literature. Such an exceptional experimental observation can be well reproduced by a theoretical model affirmatively unveiling the vital contributions from the exchange striction, while the sign difference of magnetocrystalline anisotropy can also be reasonably figured out.

---

\# These authors contributed equally to this work.

\* Corresponding authors: C.L.L. (cllu@hust.edu.cn), S.D. (sdong@seu.edu.cn).

Magnetoelectric (ME) effects refer to the control of magnetization ($M$) by electric field $E$ or control of polarization ($P$) by magnetic field $H$, which provide fascinating functionalities to develop next-generation electronic devices, e.g. ultra-sensitive sensors and energy-saving memories [1-3]. There are various manifestations of ME effects, which can be realized in many kinds of materials including but not limited to ME composites, multiferroics, as well as axion insulators [4]. Of particular interest is the most canonical linear ME effect, i.e. $H$ induces electric polarization $P_i=\alpha_{ij}H_j$, or $E$ causes a variation of magnetization $\mu_0M_j=\alpha_{ij}E_i$, with $\alpha_{ij}$ the ME tensor and $i/j$ the component of spatial coordinates [5,6]. In those linear magnetoelectrics, the interplay of broken space-inversion and time-reversal symmetries can yield a bunch of emergent physics and functions, including ferrotoroidicity, ME monopole, non-reciprocal transport of quasi-particles, etc [7-9]. Recently, it was found that the writing threshold could be reduced drastically in memory device made of the paradigmatic linear ME $Cr_2O_3$, implying the technological importance of linear ME in enhancing energy efficiency [10]. Although the studies on linear ME can be traced back to 1960s, generally the intrinsic linear magnetoelectricity to date remains weak, with typical $\alpha$ values of 1~10 ps/m, which has been a long-term issue of ME activities [5,6].

Thermodynamically it is well known that $\alpha_{ij}$ is just bounded by the permittivity $\varepsilon_{ii}$ and permeability $\mu_{jj}$ of solids, i.e. $\alpha_{ij}^2<\varepsilon_{ii}\mu_{jj}$ [11]. To obtain large $\alpha_{ij}$, nonlinear regions of $\varepsilon_{ii}$ and $\mu_{jj}$ are usually employed, especially near critical point of a ME system where the permittivity and/or permeability may be significantly amplified. For example, in those type-II multiferroics where spontaneous $P$'s are generated by some special magnetic orders, magnetic transitions or metamagnetic transitions can trigger remarkable changes of $P$'s and result in very large ME coefficients [12]. However, such a scenario can only work in the proximate region near the transition point. In this sense, the operating window of potential ME devices based on these materials would be narrow.

In fact, the integration upon $H$ is the total change of $P$ (i.e. $\Delta P$). With fixed $\Delta P$, the larger $\alpha$, the narrower operating $H$-window. Such a trade-off between ME sensitivity and operating window makes it difficult to obtain an optimal system with superior performances in both sides (large $\alpha$ & wide $H$-window), once the total $P$ generated by magnetism is small. Therefore, alternative routes should be developed.

It has been well recognized that the symmetric exchange striction allows much larger $P$'s (e.g. on the order of ~1 μC/cm$^2$ in E-type antiferromagnetic (AFM) manganites) [13,14], while other mechanisms for magnetism-driven polarization (e.g. the inverse Dzyaloshinskii-Moriya interaction (DMI) and the spin-dependent $p$-$d$ hybridization) rely on the spin-orbit coupling (SOC) and typically lead to small $P$'s (~0.001-0.1 μC/cm$^2$) [15-17]. Nevertheless, only large $P$'s from exchange striction is not sufficient. In fact, some magnetic textures (e.g. E-type phase in o-LuMnO$_3$ [18] and o-YMnO$_3$ [19]) are robust against $H$, leading to weak ME response. In other cases, large nonlinear ME effects, instead of linear one, occur in relative narrow $H$-window, triggered by the first-order phase transitions [20-22]. These facts indicate that additional physical ingredient has to be involved to obtain large linear ME.

In this Letter, the ME response of a polar magnet Fe$_2$Mo$_3$O$_8$ is studied. Our experimental study finds a colossal linear ME effect in a broad $H$-window, far beyond the consequence near the first-order transition point. Specifically, the diagonal linear ME component in its ferrimagnetic (FRM) phase is unprecedentedly large: $α_{33}$~480 ps/m at 20 K, orders of magnitude larger than reported values in literature [5]. Our theoretical calculations suggest that the exchange striction tuned polarity and sign difference of magnetocrystalline anisotropy are essential in physics.

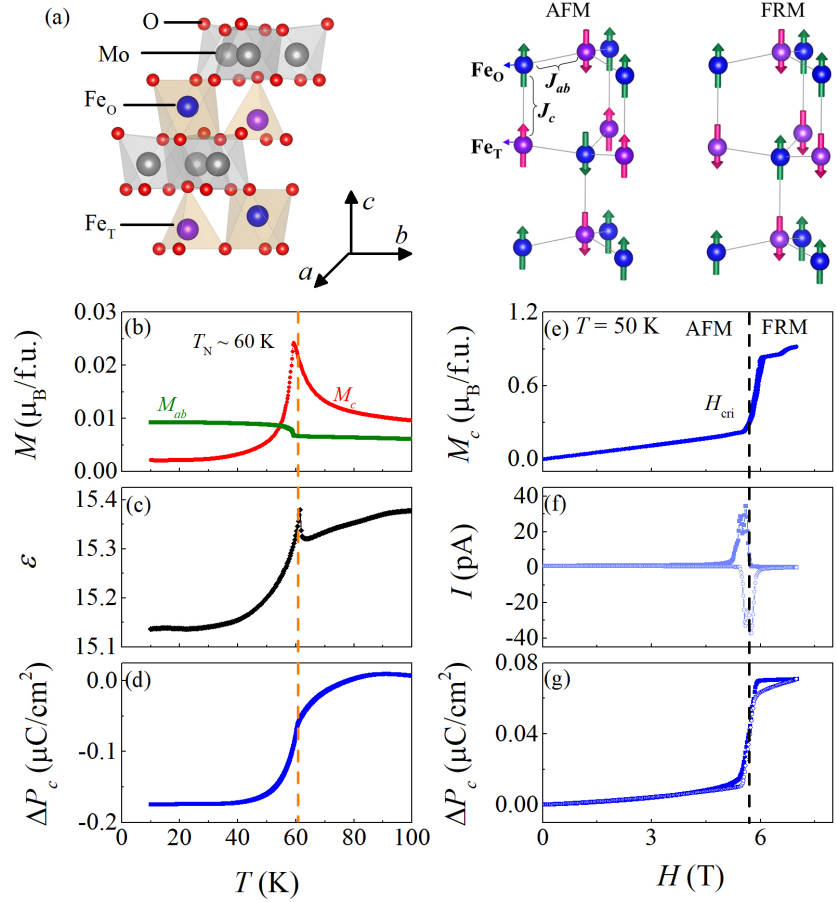

**Figure 1.** (a) Schematic of crystal and magnetic structures of $Fe_2Mo_3O_8$. (b-d) show $T$-dependence of magnetization along the $c$-axis and the $ab$-plane, $c$-axis dielectric constant $\varepsilon$, and $\Delta P_c$ along the $c$-axis, respectively. (e-g) $H$-dependences of $M_c$, magneto-current $I$, and integrated $\Delta P_c$ along the $c$-axis at $T$=50 K, respectively.

As shown in Fig. 1(a), $Fe_2Mo_3O_8$ consists of Fe-O and Mo-O blocks stacking along the $c$-axis, and the polar arrangement of $FeO_4$ tetrahedra manifests the crystallographic polarity [23,24]. The corner-shared $FeO_4$ tetrahedra and $FeO_6$ octahedra form a buckled honeycomb structure within the $ab$-plane. $Fe_2Mo_3O_8$ has an AFM ground state (magnetic point group $6'mm'$) below $T_N$~60 K with all $Fe^{2+}$-spins pointing along the $c$-axis [23], which can be evidenced by the anomalies in the temperature ($T$) dependence of $M$ as shown in Fig. 1(b). The remarkable difference between $M_c(T)$ and $M_{ab}(T)$ persists up to room temperature, implying its strong magnetic anisotropy. Corresponding to the AFM transition, a sharp peak arises in the $T$-dependence of dielectric constant $\varepsilon(T)$ (Fig. 1(c)), as well as an obvious change of $P_c(T)$ (Fig. 1(d)), revealing the onset of magnetically driven $P$ based on the polar structure. The exchange striction mechanism is responsible for this emergent $P$ [23,24].

By applying $H//c$-axis above a critical field $H_{cri}$, a FRM phase is stabilized where the Fe tetrahedral and octahedral irons ($Fe_T$ vs $Fe_O$) form individual ferromagnetic sublattices and align in opposite directions (Fig. 1(a)). As shown in Fig. 1(e), a jump of $\Delta M_c \sim 0.6$ $\mu_B$/f.u. at $H_{cri}$ indicates that $Fe_T$ and $Fe_O$ sites have different magnetic moments. Concomitantly, the AFM-FRM transition causes striking magneto-current at $H_{cri}$, and the integrated $\Delta P_c(H)$ curve exhibits a step-like anomaly with $\Delta P_c \sim 0.06$ $\mu C/cm^2$, as shown in Figs. 1(f-g), consistent with the previous report [24]. The $H$-stabilized FRM phase belongs to the magnetic point group *6m'm'*, which allows diagonal linear ME components. Previous works studied the ME response in a relatively narrow $H$-$T$ window, and indeed observed approximatively linear ME with $\alpha_{33} \sim 16$ ps/m in the FRM phase of $Fe_2Mo_3O_8$, although there were also nonlinear terms mixed-in [24,25]. A single-site mechanism, i.e. *g*-factor effect, was proposed to explain this linear ME, instead of the symmetric exchange responsible for $P$, i.e. a two-site mechanism.

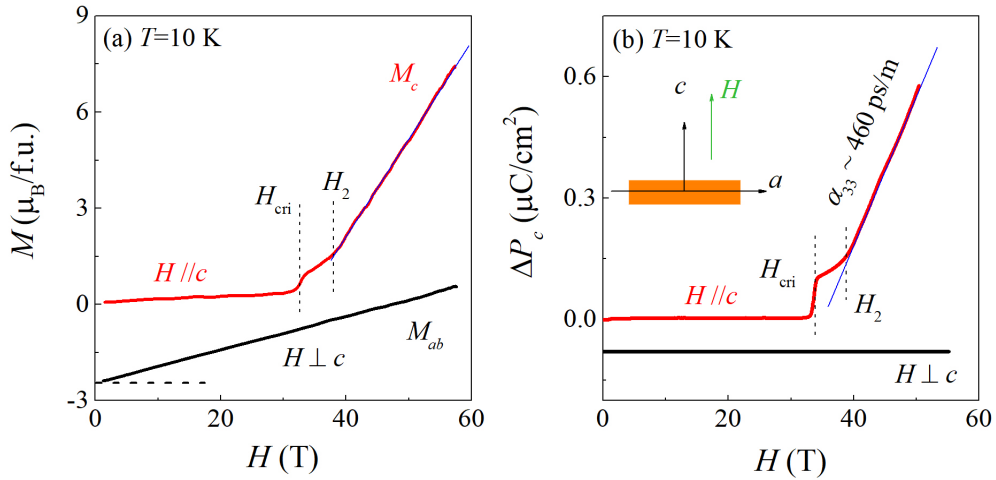

**Figure 2.** (a) Out-of-plane ($M_c$) and in-plane magnetization ($M_a$) as a function of $H$ measured at $T=10$ K. (b) $H$-dependence of $\Delta P_c$ measured with $H//c$-axis and $H\perp c$-axis at $T=10$ K. For better view, the black curves are vertically shifted from the zero point.

In the following, we performed extensive measurements of magnetization and electric polarization over a large $H$ range up to ~58 T using a pulsed high magnetic field apparatus, which allows in-depth exploration of the linear ME of the FRM phase. As shown in Fig. 2(a), $M_c(H)$ measured with $H//c$-axis at $T=10$ K also displays a clear AFM-FRM transition with a very high critical field $H_{cri} \sim 32$ T, and the related jump remains $\Delta M_c \sim 0.6$ $\mu_B$/f.u.. Slightly above $H_{cri}$, $M_c(H)$ shows another anomaly at $H_2 \sim 39$ T, and then a subsequent linear increase as $H > H_2$. The anomaly at $H_2$ will be discussed later. $M_c$ reaches $\sim 7.5$ $\mu_B$/f.u. at $H \sim 58$ T, close to the fully polarized $M \sim 8.5$ $\mu_B$/f.u. for $Fe_2Mo_3O_8$. Different from $M_c(H)$, the in-plane magnetization $M_{ab}$ continuously increases with $H$, without any anomaly up to $H \sim 58$ T, confirming the strong magnetic anisotropy of $Fe_2Mo_3O_8$.

In accordance with $M_c(H)$, $\Delta P_c(H)$ measured with $H//c$-axis at 10 K also shows an abrupt increase at $H_{cri}$, and then a crossover anomaly at $H \sim H_2$. A fascinating behavior is the emergence of remarkable and perfect linear enhancement of $\Delta P_c(H)$ above $H_2$. The estimated linear ME coefficient in a wide range is as large as 460 ps/m, far beyond other known linear ME materials. For instance, $\alpha \sim 4$ ps/m for $Cr_2O_3$ [26], $\alpha \sim 2$ ps/m for $LiNiPO_4$ [27], $\alpha \sim 10$ ps/m for $GaFeO_3$ [28], $\alpha \sim 30$ ps/m for $TbPO_4$ [29], and $\alpha \sim 40$ ps/m for $Co_4Nb_2O_9$ [30]. With the measured dielectric constant $\varepsilon \sim 15$ at $T < T_N$ for $Fe_2Mo_3O_8$ (Fig. 1(c)), one may translate $dP/dH \sim 460$ ps/m to $dE/dH \sim 2800$ mV/cm-Oe. This value is already comparable with many composite systems, such as commercially used composite $Pb(Zr,Ti)O_3$/terfrnol-D (4800 mV/cm-Oe) [31]. For $H //ab$-plane, $\Delta P_c(H)$ remains zero till the maximum field 58 T, indicating the absence of non-diagonal ME terms.

An interesting phenomenon is that both $\Delta P_c(H)$ and $M_c(H)$ show similar linear behavior above $H_2$. In particular, $M_c(H)$ exhibits considerable enhancement when $H > H_2$, suggesting drastic modification of the spin arrangement. This is distinctly different from the case of linear magnetoelectrics based on the single-site mechanisms such as the $g$-factor mechanism and single-ion anisotropy mechanism, which exhibit nearly saturated $M(H)$ behavior as appeared in $(Fe_{1-x}Zn_x)_2Mo_3O_8$ and $Ga_{2-x}Fe_xO_3$ [24,28]. Here the estimated linear $\Delta P_c(H)$ reaches $\sim 0.45$ $\mu C/cm^2$, which is much larger than the induced-$P$ due to the inverse DMI ($\sim 0.01$ $\mu C/cm^2$) [2,15]. Therefore, this remarkable linear variation $\Delta P_c(H)$ is more like to be due to the exchange striction effect (i.e. two-site mechanism). As mentioned above, complex ME response was reported at relatively low $H$-range (i.e. $H_{cri} < H < H_2$) in $Fe_2Mo_3O_8$ [24]. Therefore, the crossover at $H_2$ could be ascribed to a transition from complex ME response to pure linear ME, namely from individual magnetic-site effect to two-site mechanism.

Comprehensive characterizations of $M_c(H)$ and ME response have been performed over a broad $T$-range from 1.6 K to 50 K. As shown in Fig. 3(a), the AFM-FRM transition is well visible at all measured temperatures, and $H_{cri}$ shifts toward low field with increasing $T$. $H_2$ is not as well-identified as $H_{cri}$ especially in high-$T$ range (i.e. above 30 K), and just shows slight shift with $T$, implying different underlying physics. Corresponding behaviors are also seen in $\Delta P_c(H)$ curves measured at various $T$'s, as shown in Fig. 3(b). Importantly, both $M_c(H)$ and $\Delta P_c(H)$ display almost perfect linear behavior above $H_2$ at all $T$'s, and the slope shows clear $T$-dependence.

Although here pulsed magnetic field is used, all above ME results should be considered as static properties instead of dynamic ones, as confirmed by a comparison study of some key parameters obtained using steady and pulsed magnetic fields (see Supplementary Materials Fig. S3 [32]). The duration time $t$~10.5 ms of the $H$-pulse remains far longer than the time scale of spin dynamics which should be in the THz level for $Fe_2Mo_3O_8$ [33]. Based on the linear behaviors of both $M_c(H)$ and $\Delta P_c(H)$ at high-$H$, it is able to precisely estimate the relationship between $M_c$ and $P_c$, i.e. d$P$/d$M$. As shown in Fig. 3(c), d$P$/d$M$ distributes roughly at 0.12-0.14 below $T_N$~60 K. The obtained linear ME coefficients $\alpha_{33}$ ($H//c$) are plotted as a function of $T$ in Fig. 3(d), in which $\alpha_{33}$ reaches a plateau of ~480 ps/m below 20 K. The colossal linear magnetoelectricity can also be obtained in Zn-substituted $Fe_2Mo_3O_8$ with reduced $H_{cri}$ but robust $H_2$ (see Supplementary Materials [32]).

A previous theoretical work has studied the ME related to phase transitions of $Fe_2Mo_3O_8$ [25]. To understand the colossal linear ME in the FRM phase, we performed density functional theory (DFT) calculations. The AFM ground state and electronic band gap (~1.37 eV) can be obtained when a proper $U_{eff}$ and SOC are incorporated simultaneously, in agreement with previous theoretical works [34,35]. The FRM state is slightly higher in energy (0.14 meV/f.u.) than the AFM one. More details of the DFT calculations can be found in Supplementary Materials [32].

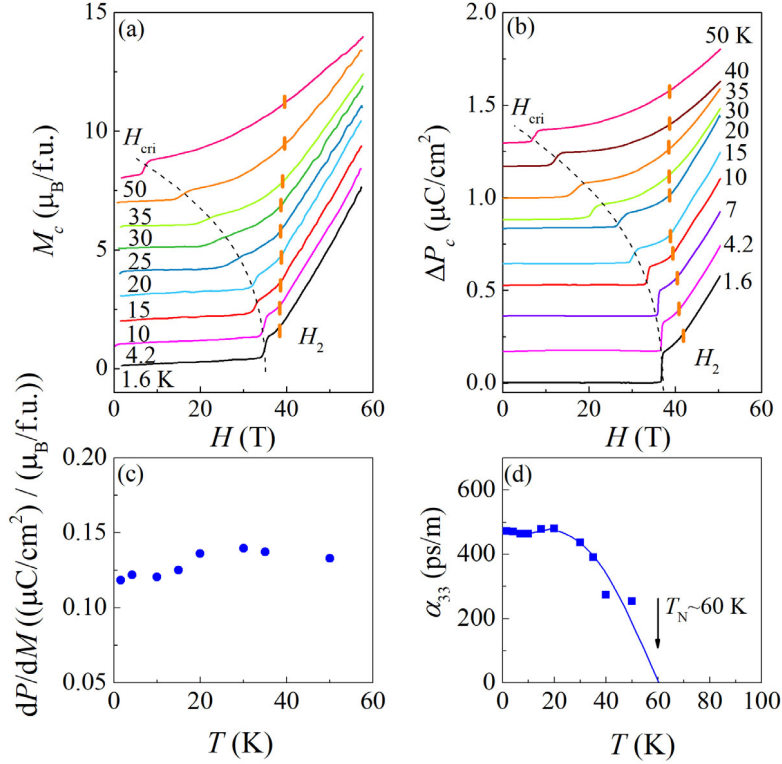

**Figure 3.** (a) and (b) show $M_c(H)$ and $\Delta P_c(H)$ at various temperatures, respectively. For better view, the curves have been shifted vertically. (c) d$P$/d$M$ as a function of temperature. (d) $T$-dependence of linear magnetoelectricity $\alpha_{33}$.

In DFT calculations, the preferred global spin direction is indeed along the $c$-aixs, while other orientations lead to higher energy. The local spin moments at Fe$_O$ and Fe$_T$ are -3.74 $\mu_B$ and 3.68 $\mu_B$ respectively, implying the high-spin states. The orbital moment at the Fe$_O$ site is almost quenched, but the SOC effect yields a significant orbital moment of 0.34 $\mu_B$ for Fe$_T$, which is parallel to its spin moment. Similar result was also predicted by earlier studies [36,37]. Noting that here the DFT moments, no matter from spin or orbital, are integrated values within the default Wigner-Seitz sphere assigned in pseudopotentials, while the corresponding integrity values of local spin and orbital moments should be 4 $\mu_B$ and 0.5 $\mu_B$ respectively [36]. Therefore, the large orbital moment of Fe$_T$ is the major contribution to the uncompensated $M$ of FRM phase, leading to the jump ~0.5 $\mu_B$/f.u. in $M_c(H)$ curve at $H_{cri}$ (Fig. 1(e)).

Because of SOC, Fe$_T$ has a much higher magnetocrystalline anisotropy energy (MAE) than Fe$_O$. According to our DFT study, the easy axis of Fe$_T$ is along the $c$-axis, with a large

MAE coefficient $A_T$=-5.11 meV/Fe. However, unexpectedly, the *c*-axis is a hard axis for Fe$_O$, with a weaker MAE $A_O$=1.97 meV/Fe. Since $A_T$ is much larger than $A_O$, the overall MAE prefers the easy axis along the *c*-axis, consistent with the neutron experiments [38]. However, this sign difference of MAE between Fe$_T$ and Fe$_O$ is beyond previous scenario [23,24], which plays a vital role to determine the ME behavior under field (to be explained below).

Under a magnetic field along the *c*-axis, the magnetic state first transits from the AFM to FRM phase at $H_{cri}$, with a net magnetization ~0.5 $\mu_B$/f.u. along the *c*-axis. In other words, the magnetic moment of Fe$_T$ (Fe$_O$) is parallel (anti-parallel) to the magnetic field (i.e. *c*-axis). Further increasing of magnetic field can only lead to the magnetic canting of Fe$_O$. Then the free energy during this spin rotation process can be expressed as:

$$F=F_0+[3J_{ab}+J_c]\cdot\cos\theta +A_O\cdot\cos^2\theta - H\cdot m_O\cdot\cos\theta, \qquad (1)$$

where $J_{ab}$ ($J_c$) is the Heisenberg-type effective exchange between in-plane (out-of-plane) nearest-neighboring Fe$_O$ and Fe$_T$; $\theta$ is the moment angle of Fe$_O$ as indicated in the inset of Fig. 4(a). The third item is the MAE of Fe$_O$; and the last item is the Zeeman energy of Fe$_O$. Here the MAE and Zeeman energy of Fe$_T$ is not explicitly expressed since they are not changing during the process and thus can be absorbed into the free energy base $F_0$.

According to Eq. 1, the starting magnetic field to rotate the Fe$_O$ moment is $[3J_{ab}+J_c-2A_O]/m_O$, corresponding to the experimental $H_2$. For comparison, $H_{cri}$ at *T*=0 K to drive the AFM to FMR transition can be expressed as $-2J_c/[m_T-m_O]$, which are not identical to $H_2$ in physics. Based on the experimental value $H_{cri}$=32 T at 1.6 K and $m_T-m_O$=0.5 $\mu_B$, $J_c$ can be derived as -0.46 meV. And our model based on Eq. 1 (with $3J_{ab}+J_c$=11.5 meV and $A_O$=1.45 meV) can well mimic the *M-H* behavior in the whole range, as shown in Fig. 4(a). First, below $H_{cri}$, the ground state is AFM, with *M*=0; Then between $H_{cri}$ and $H_2$, the magnetic state can be understood as a collinear FMR state without spin canting at *T*=0 K. Thus the $M_c$-*H* curve shows a terrace behavior between $H_{cri}$ and $H_2$ at low temperature, and thus the ME coefficient in this region is very limited, as revealed in previous studies [24]. After $H_2$, the magnetic moment of Fe$_O$ starts to rotate, resulting in the increasing of $M_c$. According to Eq. 1, the $M_c$-*H* curve is indeed linear with a slope $m_O^2/2A_O$, reaching ~7.5 $\mu_B$/f.u. at 58 T. Note the three independent coefficients ($J_c$, $3J_{ab}+J_c$, $A_O$) used in our model are very close to the DFT extracted ones (-0.07 meV, 10.55 meV, 1.97 meV). Considering the simplified spin model and tolerance of DFT values, such a consistent between model and DFT values are rather well, at least in a semiquantitative level. In particular, the sign difference of $A_T$ and $A_O$ is crucial to obtain the linear ME behavior after the terrace. Or if $A_O<0$, the magnetic transition will be in a sudden jump manner at a critical field (like what occurs at $H_{cri}$), instead of a continuous rotation of Fe$_O$ spin in a broad region.

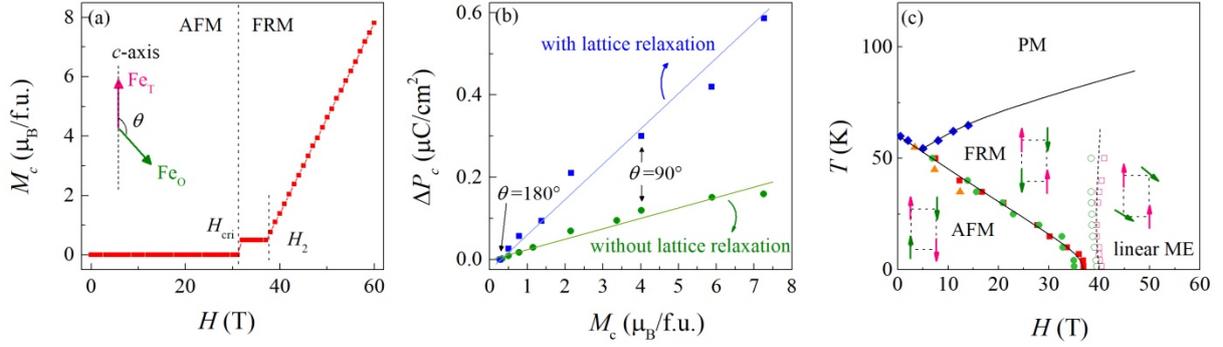

**Figure 4.** (a) The simulated $M$-$H$ evolution based on the simplified spin model (Eq. 1). Model coefficients: $m_O$=4 $\mu_B$, $J_c$=-0.46 meV, $3J_{ab}+J_c$=11.5 meV, and $A_O$=1.45 meV. (b) DFT calculated $\Delta P_c$ as a function $M_c$ during the spin rotation of $m_O$. For comparison, both the non-relaxed case and relaxed one are calculated. In the non-relaxed case, the crystalline structure is fixed as the optimized FRM phase without spin canting. Thus, only the bias of electron cloud is considered, namely pure electronic contribution. In the relaxed one, the structure is optimized with SOC enabled to mimic the full contributions from both electronic and lattice. For both cases, the ME behaviors ($\Delta P_c$-$M_c$) are linear. (c) The ME phase diagram of Fe$_2$Mo$_3$O$_8$ constructed based on experimental and theoretical data. PM: paramagnetic.

The broad-range linear ME behavior suggests that here the spin-related polarization is mainly contributed by the exchange striction, namely in the form of $\Delta P_c \sim S_{Fe} \cdot S_O \sim \cos\theta \sim \Delta M_c$, instead of the inverse DMI (in the form of $S_{Fe} \times S_O \sim \sin\theta \sim \Delta M_{ab}$). Since exchange striction is a non-relativistic effect which is relative strong, the ME coefficient here can be so large. Then the concrete value of $\Delta P_c$ can be obtained in our DFT calculation by rotating the spin of Fe$_O$, as shown in Fig. 4(b). Impressively, the $\Delta P_c$-$M_c$ curve due to the Fe$_O$ spin rotation is indeed linear. The DFT estimated slop of $\Delta P_c(M_c)$ is $dP/dM \sim 0.086$ at $T$=0 K, close to the experimental value of $dP/dM \sim 0.12$ shown in Fig. 3(c).

Based on above experimental results and theoretical calculations, a ($H$, $T$) phase diagram is summarized in Fig. 4(c), where various magnetic phase transitions and crossover behavior of the linear ME are presented. In low-$H$ region (i.e. below $H_{cri}$), the AFM state with spins aligned along the $c$-axis is stabilized, and no linear ME is identified. At $H_{cri}<H<H_2$, the AFM phase is transformed to the FRM one, and thus exhibits linear ME. Nevertheless, the linear ME appearing at $H_{cri}<H<H_2$ hosts an individual-site mechanism [24], different from the origin of the magnetically driven $P$. Above $H_2$, $H$ is sufficiently large to tune the exchange striction of neighboring $Fe^{2+}$, leading to another linear ME behavior with colossal coefficient, i.e. $\alpha_{33}$=480 ps/m at 20 K.

**Acknowledgements:** This work is supported by the National Nature Science Foundation of China (Grant Nos. 12174128, 11834002, 11704139, 92163210, and 12074135), Hubei Province Natural Science Foundation of China (Grant Nos. 2020CFA083 and 2021CFB574), the Interdisciplinary program of Wuhan National High Magnetic Field Center (WHMFC) at Huazhong University of Science and Technology (Grant No. WHMFC202205). The high magnetic field measurements were performed at WHMFC.